\documentclass[prd,twocolumn,showpacs,amsmath,amssymb,superscriptaddress,preprintnumbers,nofootinbib,showkeys]{revtex4}
\usepackage{hyperref}

\begin{document}

\newcommand{\nablab}{{\mathop {\rule{0pt}{0pt}{\nabla}}\limits^{\bot}}\rule{0pt}{0pt}}

\title{Interaction of the axionic dark matter, dynamic aether, spinor and gravity fields \\ as an origin of oscillations of the fermion effective mass}

\author{Alexander B. Balakin}
\email{Alexander.Balakin@kpfu.ru} \affiliation{Department of
General Relativity and Gravitation, Institute of Physics, Kazan
Federal University, Kremlevskaya str. 16a, Kazan 420008, Russia}

\author{Anna O. Efremova}
\email{anna.efremova131@yandex.ru} \affiliation{Department of
General Relativity and Gravitation, Institute of Physics, Kazan
Federal University, Kremlevskaya str. 16a, Kazan 420008, Russia}

\date{\today}

\begin{abstract}
In the framework of the Einstein-Dirac-axion-aether theory we consider the quartet of self-interacting cosmic fields, which includes the dynamic aether, presented by the unit timelike vector field, the axionic dark mater, described by the pseudoscalar field, the spinor field associated with fermion particles, and the gravity field. The key, associated with the mechanism of self-interaction, is installed into the modified periodic potential of the pseudoscalar (axion) field constructed on the base of a guiding function, which depends on one invariant, one pseudo-invariant and two cross-invariants containing the spinor and vector fields. The total system of the field equations related to the isotropic homogeneous cosmological model is solved; we have found the exact solutions for the guiding function for three cases: nonzero, vanishing and critical values of the cosmological constant. Based on these solutions, we obtained the expressions for the effective mass of spinor particles, interacting with the axionic dark matter and dynamic aether. This effective mass is shown to bear imprints of the cosmological epoch and of the state of the cosmic dark fluid in that epoch.
\end{abstract}
\maketitle

\section{Introduction}

The Einstein-Dirac-axion-aether theory is very effective instrument for bringing together various trends in modern Cosmology, Astrophysics and High Energy Physics. First of all, this theory deals with the Dirac spinor fields, and thus it can be applied to description of wide class of phenomena, which  occur in the fermion systems. The Dirac spinor fields, according to the classification adopted in the Standard model, describe the Dirac fermions, i.e., massive particles with a half-integer spin, which do not coincide with their antiparticles. This subclass of fermions contains, in particular, the baryons (e.g., protons and neutrons), the leptons (e.g., electrons, positrons, massive neutrinos). It is especially interesting, when one deals with the supernova and solar neutrinos, with neutrino oscillations and the problem of neutrino masses (see, e.g., \cite{book1,book2} for details). When one deals with the massless fermions, the Weyl and Majorana equations are used, as specific particular cases of the Dirac equations.

The second element of the theory is the axionic dark matter, the key participant of the cosmological events (see, e.g., \cite{a1,a2,a3} for references, historical motives and mathematical details). We follow the idea of the author of the review \cite{a1}, that now just the axions are considered to be the leading particle candidates to provide the phenomena attributed to the cosmic dark matter (e.g., the flat galactic rotation curves, the structure and properties of the dark matter halos, etc.). New aspects of this discussion and new argumentation can be found in the recent review \cite{Newa1}.

The dynamic aether, the third key element of the theory, was introduced in \cite{ae1,ae2,ae3}; it can produce the effects, which are usually associated with the influence of the cosmic dark energy \cite{de1,de2,de3}.

Unexpectedly interesting applications appear, when one considers the pairwise interactions, e.g., between spinors and dark matter, spinors and dark energy, dark matter and aether, etc. For instance, the enigma of the neutrino masses pushed to the study of coupling between neutrinos and dark matter, provoking an extensive series of works (see, e.g., \cite{NM1,NM2,NM3,NM4,NM5,NM6,NM7,NM8}).
The theory of coupling between spinor field and dark energy gives another example of such pairwise interaction (see, e.g., \cite{SE1,SE2,SE3,SE4,SE5,SE6}). In this sense, it is interesting to mention the new trend appeared in the theory of interactions in the Dark Sector: we mean the so-called dark spinors (see, e.g., \cite{DS1,DS2,DS3}).
Extensions of the canonic Einstein-aether theory \cite{ae2} have given many interesting results for the models of coupling between aether and scalar field (see, e.g., \cite{AeS1,AeS2}), for the models of interaction of the aether with the axion field \cite{B1,B11,B2}, and for the aether models including the electromagnetic field \cite{B3,B4}.

There are examples of investigations of the triple interactions: for instance, in the work \cite{TR1} the coupling between spinor, axion and scalar fields is studied; in the papers \cite{B5,B6} the coupling between aether, axions and photons is considered; in the works \cite{Saha1,Saha2,Saha3} the spinor and scalar fields are considered to be coupled to the fluids (perfect, viscous, magnetized). In some works the spinor field is presented in terms of the nonlinear and nonminimal formalisms (see, e.g., \cite{Saha4,Saha5}).

In the context of these numerous investigations, we have to clarify: what are the aims and frames of our work.
First of all, we have to say that many quoted results are formulated on the language of High Energy Physics, but we use the formalism of the Field Theory, to be more precise, we use the formalism of the Einstein-Dirac-axion-aether theory. In the framework of this theory we consider the quadruple  coupling of the gravitational, spinor, pseudoscalar and vector fields. In the total Lagrangian of the system the gravity field is presented by the classical Einstein-Hilbert term; in the nearest future we plan to use the frames of the Modified theories of gravity along the line proposed in the works \cite{SDO1,SDO2,SDO3,SDO4}. The spinor field $\psi$ (massive or massless) is described by the extended Dirac equations. The pseudoscalar field $\phi$ is associated with the axionic dark matter. The unit timelike vector field $U^i$ describes the velocity of the dynamic aether flow. We assume that the potential of the axion field $V(\phi,\Phi_*)$ contains the guiding function $\Phi_*$, which depends on four arguments $\Phi(S,P,\omega,\Omega)$. The first argument $S=\bar{\psi} \psi$ is the invariant constructed using the Dirac spinor;  the pseudoinvariant $P=\bar{\psi}\gamma^5 \psi$ contains the Dirac matrix $\gamma^5$; the scalar
$\omega =U_k \left(\bar{\psi} \gamma^k \psi \right)$ is the cross-invariant, which contains both spinor and vector fields; the pseudoscalar $\Omega = U_k \left(\bar{\psi} \gamma^5 \gamma^k  \psi \right)$ also has to be indicated as the cross -pseudoinvariant.

In this work, classifying the scalars, which could be the arguments of the guiding function, we omitted the scalars, obtained from the decomposition of the covariant derivative of the aether velocity four-vector $\nabla_k U^i$: the expansion scalar $\Theta=\nabla_kU^k$, the square of the acceleration four-vector $a^j=U^k\nabla_k U^j$, the squares of the symmetric shear tensor $\sigma_{jk}$ and of the skew-symmetric vorticity tensor $\omega_{jk}$ (see \cite{B1,B2} for details). We hope to extend the model, respectively, in the next work, but now we restrict our-selves by the scalars of zero order in derivative, according to the terminology of the Effective field theory \cite{EFT}. Keeping in mind this idea, we do not include into the Lagrangian the pseudoscalar $\nabla_k \phi (\bar{\psi} \gamma^k \psi)$, and the scalar $\nabla_k \phi (\bar{\psi} \gamma^k \gamma^5  \psi)$. As for the zero order cross-scalar $P \cdot \phi$ and cross-pseudoscalar $S \cdot \phi$, it seems to be not logical to include them into the specific potential of the axion field $V(\phi, \Phi_*(S,P,...))$, which we use for the description of this model.

Our final remark concerns the electromagnetic field and Yang-Mills fields. We assume here that these gauge $U(1)$ and $SU(N)$ symmetric fields are vanishing, and thus, the extended derivative of the Dirac spinor field does not contain the corresponding potentials. The problem is that there exists the aether-like representation of the gauge field \cite{Petrov1,Petrov2}, and the question arises: whether the aether velocity four-vector $U^k$ or its SU(N) generalizations $U^k_{(a)}$ (see, e.g., \cite{Color}) can play the role of the gauge potential? We assume that this question is still open, but in this work we follow the classic non-gauge invariant version of the dynamic aether.

The construction of the axion field potential $V(\phi,\Phi_*)$ admits the following twofold interpretation: on the one hand, the spinor field and the aether regulate the behavior of the axion field via this potential; on the other hand, the axionic backreaction modifies the Dirac equations for the spinor field, and the effective spinor mass matrix $M$ appears instead of the intrinsic mass $m$.  In other words, we assume that if the intrinsic mass of the spinor particle (e.g., of the neutrino) is equal to zero, $m=0$, then the coupling to the axionic dark matter produces an effective mass, which depends, e.g., on cosmological epoch and on the state of the axionic dark matter in that epoch.

The paper is organized as follows. In Section IIA we describe the action functional of the model; in Section IIB we specify the structure and properties of the  modified periodic potential of the axion field; in Section IIC based on the Lagrange formalism we derive the extended equations for the vector, axion, spinor and gravitational fields. Section III contains cosmological application of the elaborated model. In Section IIIA we reduce the field equations of the model to the symmetry associated with the isotropic homogeneous cosmology. In Section IIIB we obtain exact solutions of the complete set of the reduced field equations,  and discuss the properties of the exact solutions to the guiding function $\Phi_*$ for two cases: with and without cosmological constant,  $\Lambda \neq 0$ and $\Lambda = 0$. In Section IV we discuss the properties of the obtained effective mass attributed to the spinor field coupled to the axionic dark matter. Section V contains discussion and conclusions.

\section{The formalism of the Einstein-Dirac-aether-axion theory}

\subsection{Action functional}

The action functional describing the system of interacting gravitational, spinor, pseudoscalar (axion) vector fields can be written as follows:
$$
-{\cal S}= \int d^4x \sqrt{-g}\left\{ \frac{1}{2\kappa}\left[ R+2\Lambda + \lambda\left(g_{mn}U^mU^n - 1 \right) +
\right. \right.
$$
$$
\left. \left. +K^{ab}_{\ \ mn} \nabla_a U^m  \nabla_b U^n \right] + \frac12\Psi^2_0\left[V - \nabla^k \phi \nabla_k \phi \right] + \right.
$$
\begin{equation}
\left. + \frac{i}{2}[\bar{\psi}\gamma^{k}D_{k}\psi-D_{k}\bar{\psi}\gamma^{k}\psi]-m\bar{\psi}\psi - L_{(\rm baryon)} \right\} \,.
\label{action}
\end{equation}
As usual, $g$ is the determinant of the metric, $\kappa$ is the Einstein constant, $R$ is the Ricci scalar, $\Lambda$ is the cosmological constant, $\nabla_k$ is the covariant derivative.
$\lambda$ is the Lagrange multiplier, $U^i$ is the unit timelike vector field, which is associated with the velocity four-vector of the aether flow.
The object
\begin{equation}
K^{ab}_{\ \ mn} {=} C_1 g^{ab}g_{mn} {+} C_2 \delta^a_m \delta^b_n {+} C_3 \delta^a_n \delta^b_m {+} C_4 U^a U^b g_{mn}
\label{K}
\end{equation}
contains four phenomenological constants $C_1$, $C_2$, $C_3$, $C_4$ and presents the so-called constitutive tensor in the model of the dynamic aether \cite{ae2}.

The dimensionless pseudoscalar field $\phi$ is associated with the axions, hypothetical massive pseudo-Goldstone bosons \cite{PQ,Weinberg,Wilczek}; $V$ is the potential of the pseudoscalar field, and the parameter $\Psi_0$ is connected with the constant of the axion-photon coupling $g_{A \gamma \gamma}$ by the relationship $\frac{1}{\Psi_0}=g_{A \gamma \gamma}$.

The matrix-column $\psi$ describes the spinor field, $\bar{\psi}$ is its Dirac conjugated field, $\bar{\psi}= {\psi^{*}}^T \gamma^0$. The matrices $\gamma^k$ satisfy the relationships
\begin{equation}
\gamma^m \gamma^n   {+}  \gamma^n \gamma^m =2 E g^{mn} \,,
\label{anti}
\end{equation}
where $E$ is the unit $4 \times 4$ matrix. The matrices $\gamma^k$ are connected with the standard constant Dirac matrices $\gamma^{(a)}$ via the tetrad four-vectors $X^{m}_{(a)}$  by the convolution $\gamma^{k} = X^k_{(a)}\gamma^{(a)}$ with respect to the tetrad index $(a)$.
The constant Dirac matrices satisfy the relationships
\begin{equation}
\gamma^{(a)} \gamma^{(b)} {+} \gamma^{(b)}\gamma^{(a)} =2 E \eta^{(a)(b)}  \,,
\label{anti2}
\end{equation}
where $\eta^{(a)(b)}$ is the Minkowski metric.
The quartet of the tetrad four-vectors $X^{m}_{(a)}$ is known to satisfy two normalization - orthogonality conditions:
\begin{equation}
g_{mn}X^{m}_{(a)}X^{n}_{(b)}=\eta_{(a)(b)} \,, \quad \eta^{(a)(b)}X^{m}_{(a)}X^{n}_{(b)}=g^{mn} \,.
\label{tetrad}
\end{equation}
The term $D_k$ defines the spinor covariant derivatives, which are given by
\begin{equation}
D_{k}\psi = \partial_{k}\psi-\Gamma_{k}\psi \,, \quad
D_{k}\bar{\psi}=\partial_{k}\bar{\psi}+\bar{\psi}\Gamma_{k} \,.
\label{Fock}
\end{equation}
The Fock-Ivanenko matrices $\Gamma_{k}$ \cite{FI} can be expressed via the covariant derivatives of the tetrad four-vectors:
\begin{equation}
\Gamma_{k}=\frac{1}{4}g_{mn}X^{(a)}_{s}\gamma^{s}\gamma^{n} \nabla_{k}X^{m}_{(a)} \,.
\label{Gamma}
\end{equation}
The parameter $m$ describes the mass of the spinor field.

Finally, we would like to focus on the term $L_{(\rm baryon)}$, the last element of the action functional (\ref{action}). Formally speaking, the baryons (protons, neutrons, etc) are the fermions, the heavy spinor particles, and thus they also can be described by the Dirac equations, if we work in the paradigm of the Field Theory. Clearly, in general case in order to describe a multi-fermion system, we have to introduce the corresponding  multi-component spinors, instead of the four-component one. There exists an alternative to this approach: one can consider the baryonic matter in terms of a multi-component fluid with the stress-energy tensor of the perfect fluid. It is necessary to use this approach, e.g., when the spinor field, which we analyze, presents the massless neutrino.

\subsection{The structure of the potential of the axion field}

We assume that the modified periodic potential of the axion field is given by the function
\begin{equation}
V(\phi,\Phi_{*}) = \frac{m^2_A \Phi^2_{*}}{2\pi^2} \left[1- \cos{\left(\frac{2 \pi \phi}{\Phi_{*}}\right)} \right] \,.
\label{V}
\end{equation}
This potential is indicated as periodic, since $V(\phi {+} n \Phi_*)= V(\phi)$; it has the minima at $\phi{=}n \Phi_{*}$. Near the minima, when
$\phi \to n \Phi_{*} {+} \tilde{\phi}$, and $|\tilde{\phi}|$ is small, the potential takes the standard form $V \to m^2_A \tilde{\phi}^2$, where $m_A$ is the axion mass.
The function $\Phi_{*}$ plays the role of vacuum average value of the axion field; we assume that it is not a constant, and it depends on coordinates via some auxiliary invariants.
In the works \cite{B1,B11} these invariants were constructed using the covariant derivative of the velocity four-vector, associated with the dynamic aether flow. In the papers \cite{BG1,BG2} we have used the moduli of the Killing vectors as auxiliary invariants, the arguments of the guiding function $\Phi_*$. Now we develop this idea and assume that the guiding function   $\Phi_*$ can depend on
the spinor field via the following four invariants (two scalars and two pseudoscalars):
$$
S=\bar{\psi} \psi \,, \quad P=\bar{\psi}\gamma^5 \psi \,,
$$
\begin{equation}
\omega =U_k \left(\bar{\psi}  \gamma^k \psi \right)\,, \quad \Omega = U_k \left(\bar{\psi} \gamma^5 \gamma^k  \psi \right) \,.
\label{scalars}
\end{equation}
We use the following definition for the matrix $\gamma^5$.
\begin{equation}
\gamma^5 = \frac{1}{4!} \epsilon_{mnpq} \gamma^m \gamma^n \gamma^p \gamma^q \,,
\label{gamma5}
\end{equation}
where $\epsilon_{mnpq}$ is the Levi-Civita tensor expressed via the Levi-Civita antisymmetric symbol $E_{mnpq}$ as follows:
\begin{equation}
\epsilon_{mnpq} = \sqrt{-g} E_{mnpq} \,, \quad E_{0123} = -1 \,.
\label{Levi}
\end{equation}
The matrix $\gamma^5$ does not depend on the metric, since
\begin{equation}
\gamma^5 = \sqrt{-g} \gamma^0 \gamma^1 \gamma^2 \gamma^3 = \gamma^{(0)} \gamma^{(1)}\gamma^{(2)}\gamma^{(3)} = \gamma^{(5)} \,.
\label{2gamma5}
\end{equation}
Two last quantities in (\ref{scalars}), $\omega$ and $\Omega$, can be indicated as cross-invariant and cross - pseudoinvariant, respectively, since they contain quantities, which characterize both the aether flow and spinor field. Let us mention that in many aspects we follow the approach presented in the series of works \cite{Saha1}-\cite{Saha5}; the authors of these works have focused on the coupling of the scalar and spinor fields via the kinetic term  $\nabla^k \varphi \nabla_k \varphi \cdot F(S,P,...)$, so one can say that we extend this approach considering the axion field instead of scalar, introducing the vector field related to the dynamic aether,  and modifying the potential $V$ instead of the kinetic term.

\subsection{Model field equations}

\subsubsection{Equations for the aether field}

Variation of the action functional (\ref{action}) with respect to the Lagrange multiplier $\lambda$ and to the four-vector $U^i$ gives, respectively, the normalization condition
\begin{equation}
g_{mn}U^m U^n =1 \,,
\label{norm}
\end{equation}
and the dynamic equation for the aether flow
\begin{equation}
\nabla_a {\cal J}^{a j}  = \lambda  U^j  +  C_4 DU_m \nabla^j U^m + \kappa I^j \,.
\label{Jacobson}
\end{equation}
The tensor ${\cal J}^{a}_{ \ j}$ is standardly presented as
$$
{\cal J}^{a}_{ \ j} = K^{ab}_{\ \ jn} \nabla_b U^n =
$$
\begin{equation}
= C_1 \nabla^a U_j + C_2 \delta^a_j \Theta + C_3\nabla_j U^a + C_4 U^a DU_j  \,,
\label{J}
\end{equation}
with the convective derivative $D = U^k \nabla_k$.
The four-vector $I^j$ is given by
$$
I^j =  \frac{m^2_{A}\Psi^2_0 \Phi_{*}}{2\pi^2}  \left[ 1{-} \cos{\left(\frac{2 \pi \phi}{\Phi_{*}}\right)}
{-}\frac{\pi \phi}{\Phi_{*}}  \sin{\left(\frac{2 \pi \phi}{\Phi_{*}}\right)} \right] \times
$$
\begin{equation}
\times \left[\frac{\partial \Phi_*}{\partial \omega} \left(\bar{\psi}\gamma^j \psi \right)  {+}
\frac{\partial \Phi_*}{\partial \Omega} \left(\bar{\psi}\gamma^5 \gamma^j \psi \right) \right] \,.
\label{I}
\end{equation}
Below, keeping in mind the compactness of formulas, we use many times the function $T=T(\phi, \Phi_*)$, defined as follows:
\begin{equation}
T = \frac{m_A^2 \Psi^2_0 \Phi_*}{2\pi^2} \left[ 1{-} \cos{\left(\frac{2 \pi \phi}{\Phi_{*}}\right)} {-} \frac{\pi \phi}{\Phi_*} \sin{\left(\frac{2 \pi \phi}{\Phi_{*}}\right)} \right] \,.
\label{T}
\end{equation}
Thus the four-vector $I^j$ can be shortly rewritten as
\begin{equation}
I^j = T  \left[\frac{\partial \Phi_*}{\partial \omega} \left(\bar{\psi}\gamma^j \psi \right)  {+}
\frac{\partial \Phi_*}{\partial \Omega} \left(\bar{\psi}\gamma^5 \gamma^j \psi \right) \right] \,.
\label{IT}
\end{equation}
The function $T=T(\phi, \Phi_*)$ takes zero value, when the axions are in the equilibrium state, $\phi {=} n \Phi_*$, and the integer $n$, describing the serial number of the equilibrium level, is arbitrary.

\subsubsection{Equations for the axion field}

Variation of the action functional (\ref{action}) with respect to the axion field gives the equation
\begin{equation}
\nabla^k \nabla_k \phi  = -   \frac{m^2_A \Phi_{*}}{2\pi} \sin{\left(\frac{2 \pi \phi}{\Phi_{*}}\right)} \,.
\label{Klein}
\end{equation}
The spinor field predetermines the structure of the function $\Phi_*(S,P,H,\Omega)$, and thus it regulates the behavior of the axion field.

\subsubsection{Equations for the spinor field}

Variation of the action functional (\ref{action}) with respect to the quantities $\bar{\psi}$ and $\psi$ gives, respectively,
\begin{equation}
i\gamma^{n} D_{n}\psi - M \psi =0 \,, \quad i D_{n}\bar\psi\gamma^{n} + \bar\psi M  = 0 \,,
\label{Dirac}
\end{equation}
where the matrix $M$ is presented by the formula
\begin{equation}
M {=} m E {-} T \left[\frac{\partial \Phi_*}{\partial S} E  {+} \frac{\partial \Phi_*}{\partial P} \gamma^5 {+} \left(\frac{\partial \Phi_*}{\partial \omega} E  {+}
\frac{\partial \Phi_*}{\partial \Omega} \gamma^5 \right) U_k \gamma^k \right].
\label{M}
\end{equation}
If we consider the effective mass as the scalar function, we can use the formula
$$
<M> \equiv \frac{(\bar{\psi} M \psi)}{(\bar{\psi}\psi)} {=}
$$
\begin{equation}
= m  {-} \frac{T}{S} \left[S \frac{\partial \Phi_*}{\partial S}   {+} P \frac{\partial \Phi_*}{\partial P} {+} \omega \frac{\partial \Phi_*}{\partial \omega}  {+}
\Omega \frac{\partial \Phi_*}{\partial \Omega} \right].
\label{2M}
\end{equation}
According to (\ref{Dirac}), (\ref{M}) the axion field $\phi$ and the aether velocity form an effective mass of the spinor field.

\subsubsection{Equations for the gravity field}

Variation of the action functional (\ref{action}) with respect to the metric yields
\begin{equation}
R_{pq}{-}\frac12 g_{pq} R {-} \Lambda g_{pq} = T^{\rm U}_{pq} {+} \kappa T^{(\rm A)}_{pq} {+} \kappa T^{(\rm D)}_{pq} {+} \kappa T^{(\rm B)}_{pq} \,.
\label{Ein}
\end{equation}
The standard stress-energy tensor of the aether $T^{(\rm U)}_{pq}$ is known to have the form:
\begin{equation}
T^{(\rm U)}_{pq} =
\frac12 g_{pq} \ K^{abmn} \nabla_a U_m \nabla_b U_n{+} \lambda U_pU_q  {+}
\label{TU}
\end{equation}
$$
{+}\nabla^m \left[U_{(p}{\cal J}_{q)m} {-}
{\cal J}_{m(p}U_{q)} {-}
{\cal J}_{(pq)} U_m\right]+
$$
$$
+C_1\left[(\nabla_m U_p)(\nabla^m U_q) {-}
(\nabla_p U_m )(\nabla_q U^m) \right]
{+}C_4 D U_p D U_q  \,.
$$
The stress-energy tensor of the axion field contains one standard  and one new elements:
\begin{equation}
T^{(\rm A)}_{pq} = \Psi^2_0 \left[\nabla_p \phi \nabla_q \phi +\frac12 g_{pq}\left(V- \nabla_k \phi \nabla^k \phi \right) \right] +
\label{TA}
\end{equation}
$$
+ \frac12 T \left\{\frac{\partial \Phi_*}{\partial \omega} \left[U_p \left(\bar{\psi}\gamma_q \psi \right) + U_q \left(\bar{\psi}\gamma_p \psi \right)\right]  {+} \right.
$$
$$
\left. + \frac{\partial \Phi_*}{\partial \Omega} \left[U_p \left(\bar{\psi}\gamma^5 \gamma_q \psi\right) + U_q \left(\bar{\psi}\gamma^5 \gamma_p \psi \right) \right] \right\} \,.
$$
The new element in (\ref{TA}), which is proportional to the function $T$, is originated from the variation of the potential $V$ with respect to the  metric. Here we used the formulas
\begin{equation}
\frac{\delta X^n_{(a)}}{\delta g^{pq}} = \frac14 \left(X_{p (a)} \delta^n_q + X_{q (a)} \delta^n_p  \right) \,,
\label{varX1}
\end{equation}
\begin{equation}
\frac{\delta g_{mn}}{\delta g^{pq}} = - \frac12 \left(g_{mp}g_{nq} + g_{np}g_{mq}\right) \,,
\label{varX2}
\end{equation}
\begin{equation}
\frac{\delta \gamma^{(a)}}{\delta g^{pq}} = 0 \,, \quad \frac{\delta \gamma^5}{\delta g^{pq}} = 0 \,, \quad \frac{\delta U^j}{\delta g^{pq}} = 0 \,.
\label{varX3}
\end{equation}
Details of variation procedure for the tetrad four-vectors can be found, e.g., in \cite{X1,X2}.

The stress-energy tensor of the spinor field $T^{(\rm D)}_{pq}$ is
\begin{equation}
T^{(\rm D)}_{pq} = - g_{pq} L_{(\rm S)} +
\label{TD}
\end{equation}
$$
+ \frac{i}{4}\left[\bar\psi \gamma_{p} D_{q}\psi {+} \bar\psi \gamma_{q} D_{p}\psi {-} (D_{p}\bar\psi) \gamma_{q} \psi {-} (D_{p}\bar\psi) \gamma_{q} \psi \right] \,.
$$
The term $L_{(\rm S)}$ appeared in (\ref{TD})
\begin{equation}
L_{(\rm S)} =
\left\{\frac{i}{2}\left[\bar\psi \gamma^{k} D_{k}\psi - (D_{k}\bar\psi) \gamma^{k} \psi \right] - m \bar\psi \psi \right\} \,,
\label{LD}
\end{equation}
takes the following form on the solutions to the modified Dirac equations:
$L_{(\rm S)} = \bar\psi(M{-}m) \psi $.
Now, in contrast to the classical case, when $\Phi_*$ is constant, the effective mass matrix $M$ (see (\ref{M})) does not coincide with the intrinsic mass matrix $m E$, and the term  $L_{(\rm S)}$ does not vanish.

The last term in the right-hand side of (\ref{Ein}) describes the contribution of the baryonic matter, if it is considered as the perfect fluid:
\begin{equation}
T^{(\rm B)}_{pq} = (W+\Pi) V_p V_q - \Pi g_{pq} \,,
\label{baryon}
\end{equation}
where $W$ and $\Pi$ are the baryonic energy density and pressure, respectively, and the velocity four-vector of the fluid, $V^p$, is the unit timelike eigen-vector of the tensor $T^{(\rm B)}_{pq}$, i.e., $T^{(\rm B)}_{pq}V^p = W V_q$ (the Landau-Lifshitz definition).

\section{Application: The model of isotropic homogeneous Universe}

The isotropic homogeneous cosmological model is a good starting point for analysis of the problem of the induced fermion masses. We see that this problem can be successfully solved also in the framework of the anisotropic Bianchi models, however we focus now on the Friedmann - type model with the cosmological constant in order to clarify physical aspects of the problem.

\subsection{Reduced field equations}

\subsubsection{Metric, tetrad and connection coefficients}

For description of the isotropic homogeneous spacetime we use the  metric
\begin{equation}
ds^2 = dt^2 - a^2(t)(dx^2+dy^2+dz^2) \,,
\label{metric}
\end{equation}
the Hubble function
$H(t)= \frac{\dot{a}}{a}$, and the set of the tetrad four-vectors
$$
X^i_{(0)} = U^i = \delta^i_0 \,, \quad X^i_{(1)} = \delta^i_1 \frac{1}{a(t)} \,,
$$
\begin{equation}
X^i_{(2)} = \delta^i_2 \frac{1}{a(t)} \,, \quad X^i_{(3)} = \delta^i_3 \frac{1}{a(t)} \,.
\label{tetradF}
\end{equation}
The spinor connection coefficients $\Gamma_k$ have now very simple form
$$
\Gamma_0 = 0 \,, \quad \Gamma_1 = \frac12 \dot{a} \gamma^{(1)} \gamma^{(0)} \,,
$$
\begin{equation}
\Gamma_2 = \frac12 \dot{a} \gamma^{(2)} \gamma^{(0)} \,, \quad \Gamma_3 = \frac12 \dot{a} \gamma^{(3)} \gamma^{(0)} \,.
\label{GFriedmann}
\end{equation}
As a direct consequence of (\ref{GFriedmann}) we obtain the auxiliary formula
\begin{equation}
\gamma^{k} \Gamma_k = - \frac32 H \gamma^0 = -  \Gamma_k \gamma^{k} \,.
\label{conseq}
\end{equation}

\subsubsection{Reduced evolutionary equation for the  unit vector field}

The symmetry of the model requires that the aether velocity four-vector has to have the form $U^j = \delta^j_0$, and thus the covariant derivative is
\begin{equation}
\nabla_k U^m = H(t)\left(\delta^m_1 \delta_k^1 + \delta^m_2 \delta_k^2 + \delta^m_3 \delta_k^3 \right) \,.
\label{nablaU}
\end{equation}
The corresponding acceleration four-vector vanishes $DU^k=0$, and the expansion scalar $\Theta = \nabla_k U^k$ is equal to $\Theta = 3H$. Thus, the tensor $J^{aj}$ is now symmetric, and its components can be written as follows:
\begin{equation}
J^a_j = H \left[(C_1{+}C_3) \left(\delta^a_1 \delta_j^1 {+} \delta^a_2 \delta_j^2 {+} \delta^a_3 \delta_j^3 \right){+} 3C_2 \delta^a_j \right] \,.
\label{JFRiedmann}
\end{equation}
The equations for the unit vector field (\ref{Jacobson}) take the form
\begin{equation}
3\delta^0_j \left[C_2 \dot{H} - (C_1+C_3) H^2\right] = \lambda \delta^0_j + \kappa I_j \,.
\label{reduceJac}
\end{equation}
Clearly, due to the model isotropy, the spatial components of the four-vector $I_j$ have to be equal to zero, $I_{\alpha}{=}0$, where $\alpha {=}1,2,3$. It is possible in one of the following three cases.
In the first case we can require
\begin{equation}
\bar{\psi} \gamma^{\alpha} \psi = 0 \,, \quad  \bar{\psi} \gamma^5 \gamma^{\alpha} \psi = 0 \,,
\label{iso}
\end{equation}
and obtain six conditions for the four complex spinor components. The second case corresponds to the equilibrium state of the axionic dark matter, and the condition $\phi {=} n \Phi_*$ leads to $I_j = 0$ because of the vanishing of the function $T$ (\ref{T}). In the third case we deal with the requirements $\frac{\partial \Phi_*}{\partial \omega}=0$ and $\frac{\partial \Phi_*}{\partial \Omega}=0$, i.e., the function $\Phi_*$ depends on $S$ and $P$ only.

When $I_{\alpha}=0$, we obtain that only one equation for the velocity four-vector from the set (\ref{reduceJac}) is nontrivial
\begin{equation}
3\left[C_2 \dot{H}- (C_1+C_3)H^2 \right] = \lambda + \kappa I_0 \,.
\label{keyJac}
\end{equation}
This equation gives the Lagrange multiplier
\begin{equation}
 \lambda(t) {=} 3\left[C_2 \dot{H}{-} (C_1{+}C_3)H^2 \right]  {-} \kappa T \left[\omega \frac{\partial \Phi_*}{\partial \omega} {+}
\Omega \frac{\partial \Phi_*}{\partial \Omega}  \right]   \,.
\label{lambda}
\end{equation}
This quantity contributes to the gravity field equation via $T^{(\rm U)}_{ik}$ (see (\ref{TU})).

\subsubsection{Reduced evolutionary equation for the axion field}

Field equation (\ref{Klein}) takes now the form
\begin{equation}
\ddot{\phi} + 3 H \dot{\phi} = -   \frac{m^2_A \Phi_{*}}{2\pi} \sin{\left(\frac{2 \pi \phi}{\Phi_{*}}\right)} \,.
\label{ddotKlein}
\end{equation}
Generally, this nonlinear equation with coefficients depending on time can be analyzed only numerically, but there are special cases, which we discuss below.

 \subsubsection{Reduced evolutionary equation for the spinor field }

The symmetry of the problem hints, that one can search for the components of the spinor field as the functions of time only; then the Dirac equations simplify essentially
\begin{equation}
i \gamma^{0}\left(\partial_0 + \frac32 H \right) \psi = M \psi \,,
\label{DF1}
\end{equation}
\begin{equation}
i \left(\partial_0 + \frac32 H \right) \bar{\psi} \gamma^0 = - \bar{\psi} M \,.
\label{DF2}
\end{equation}
The replacement
\begin{equation}
\psi = a^{-\frac32} \Psi \,, \quad \bar{\psi} = a^{-\frac32} \bar{\Psi}
\label{reopace}
\end{equation}
reduces the Dirac equations, yielding
\begin{equation}
i\gamma^{(0)} \dot{\Psi} = M \Psi \,, \quad i \dot{\bar{\Psi}} \gamma^{(0)} = - \bar{\Psi} M  \,,
\label{keyD}
\end{equation}
where the dot denotes the derivative with respect to time. Let us consider the consequences of these equations.

\subsubsection{Evolution of the spinor invariants}

Keeping in mind (\ref{keyD}), we can calculate the rate of evolution of the invariants $S$, $P$, $\omega$ and $\Omega$.
Let us demonstrate the method of derivation on the example of the scalar $S$. First, we see that
$$
\frac{d}{dt}\left(\bar{\psi} \psi \right) =
\frac{d}{dt}\left(\bar{\Psi} a^{-3}\Psi \right) =
$$
$$
= {-}3H \left(\bar{\psi} \psi \right) {+} a^{-3}\left[\left(\frac{d}{dt}\bar{\Psi}\right) {\gamma^{(0)}}^2 \Psi {+} \bar{\Psi}{\gamma^{(0)}}^2 \left(\frac{d}{dt} \Psi\right) \right]{=}
$$
\begin{equation}
= {-}3H \left(\bar{\psi} \psi \right) {+} i \bar{\psi} \left(M\gamma^0 {-} \gamma^0 M \right)\psi \,.
\label{dotS1}
\end{equation}
Using the formula for the effective mass matrix $M$ (\ref{M}) with $T$ given by (\ref{T})
we can rewrite the evolutionary equation for the scalar $S$ as follows
\begin{equation}
\dot{S} + 3H S = - 2 i T \left(\Omega \frac{\partial \Phi_*}{\partial P} + P \frac{\partial \Phi_*}{\partial \Omega} \right) \,.
\label{dotS2}
\end{equation}
Similarly, for the pseudoinvariant $P$ we can write the evolutionary equation
\begin{equation}
\dot{P} + 3H P =  i \bar{\psi} \left(M \gamma^0 \gamma^5 - \gamma^5 \gamma^0 M \right)\psi \,,
\label{dotP}
\end{equation}
from which we obtain
\begin{equation}
\dot{P} + 3H P =-2im \Omega + 2 i T \left(\Omega \frac{\partial \Phi_*}{\partial S} + S \frac{\partial \Phi_*}{\partial \Omega} \right) \,.
\label{dotP2}
\end{equation}
For the  scalar $\omega$ and pseudoscalar $\Omega$ the results are
\begin{equation}
\dot{\omega} + 3H \omega = 0 \,,
\label{dotomega}
\end{equation}
\begin{equation}
\dot{\Omega} + 3H \Omega = -i \bar{\psi} \left(M \gamma^5 + \gamma^5 M \right)\psi \,,
\label{dotOmega}
\end{equation}
or equivalently
\begin{equation}
\dot{\Omega} + 3H \Omega  = -2im P + 2 i T \left(P \frac{\partial \Phi_*}{\partial S} - S \frac{\partial \Phi_*}{\partial P} \right) \,.
\label{dotOmega2}
\end{equation}
When the axionic dark matter is in the equilibrium state, $\phi = n \Phi_*$, and consequently,  $T(\phi, \Phi_*)=0$, the evolutionary equations (\ref{dotS2}) - (\ref{dotOmega2}) convert into
\begin{equation}
\dot{S} + 3H S = 0 \,, \quad \dot{\omega} + 3H \omega = 0 \,,
\label{good1}
\end{equation}
\begin{equation}
\dot{P} + 3H P = -2im \Omega \,, \quad \dot{\Omega} + 3H \Omega = -2im P \,.
\label{good2}
\end{equation}
Then we obtain the following exact solutions to these equations
\begin{equation}
S(t)= S(t_0) \left(\frac{a(t_0)}{a(t)} \right)^3 \,,
\label{good3}
\end{equation}
\begin{equation}
\omega(t)= \omega(t_0) \left(\frac{a(t_0)}{a(t)} \right)^3 \,,
\label{good4}
\end{equation}
\begin{equation}
P(t){=} \frac{a^3(t_0)}{a^3(t)} \left[P(t_0) \cos{2m (t{-}t_0)} {+} \Omega(t_0) \sin{2m (t{-}t_0)} \right],
\label{good5}
\end{equation}
\begin{equation}
\Omega(t){=} \frac{a^3(t_0)}{a^3(t)} \left[\Omega(t_0) \cos{2m (t{-}t_0)} {-} P(t_0) \sin{2m (t{-}t_0)} \right].
\label{good6}
\end{equation}
\begin{equation}
P^2(t)+ \Omega^2(t){=} \left[P^2(t_0)+ \Omega^2(t_0)\right] \frac{a^6(t_0)}{a^6(t)} \,.
\label{good7}
\end{equation}
Below we use this finding for reconstruction of the guiding function $\Phi_*$.

\subsubsection{Key equation for the gravity field}

We work now in the approximation, for which the cosmological baryonic matter is pressureless, $\Pi=0$, the fluid velocity coincides with the aether velocity, $U^i=V^i$, and the stress-energy tensor (\ref{baryon}) is divergence-free, i.e., $\nabla^p T^{(\rm B)}_{pq}=0$.
These requirements give us the known relationship for the cosmic dust $W(t)=W(t_0) \left[\frac{a(t_0)}{a(t)}\right]^3$.
Taking into account the structure of the stress-energy tensors of the unit vector, axion and spinor fields given by (\ref{TU}), (\ref{TA}), (\ref{TD}), respectively, we find that the key equation for the gravity field reads
$$
3H^2 \Gamma {-} \Lambda  {-} \kappa \left[m S(t_0){+}W(t_0)\right] \left(\frac{a(t_0)}{a(t)}\right)^3  =
$$
\begin{equation}
= \frac12 \kappa \Psi^2_0 \left\{\frac{m^2_A \Phi^2_{*}}{2\pi^2} \left[1{-} \cos{\left(\frac{2 \pi \phi}{\Phi_{*}}\right)} \right]{+} {\dot{\phi}}^2  \right\} \,,
\label{keyGravity}
\end{equation}
where we introduced a new auxiliary parameter:
\begin{equation}
\Gamma = 1 + \frac12 (C_1+3C_2+C_3) \,.
\label{GammaC}
\end{equation}
Other Einstein's equations are the differential consequences of the evolutionary equations for the aether, axion and spinor fields.

\subsection{Equilibrium axionic dark matter: \\ How do we search for the guiding function $\Phi_*$? }

\subsubsection{Key equation for the guiding function $\Phi_*$}

We indicate the state of the axionic dark matter as an Equilibrium state, when the potential $V(\phi,\Phi_*)$ and its derivative $\frac{\partial V}{\partial \phi}$ take zero values. For the periodic potential (\ref{V}) these conditions are satisfied, if the pseudoscalar field is in one of the minima of the potential, i.e., $\phi = n \Phi_*$, where $n$ is an integer.
For the Equilibrium state the equation (\ref{ddotKlein}) converts into the equation for the guiding function $\Phi_*(t)$:
\begin{equation}
\ddot{\Phi}_* + 3 H \dot{\Phi}_* =0 \,.
\label{ddotPhi}
\end{equation}
This equation does not depend on the integer $n$ and admits the first integral
\begin{equation}
 \dot{\Phi}_{*}(t) = \dot{\Phi}_*(t_0) \left(\frac{a(t_0)}{a(t)}\right)^3 \,.
\label{dotPhi}
\end{equation}
Below we use the auxiliary variable $x$ defined as follows:
\begin{equation}
x= \frac{a(t)}{a(t_0)} \,,  \quad \frac{d}{dt} = xH(x)\frac{d}{dx} \,.
\label{v}
\end{equation}
In terms of this new variable the Hubble function extracted from (\ref{keyGravity}) at $\phi {=} \Phi_*$ (we choose the level $n=1$ as the basic one) is given by
\begin{equation}
H(x) = \sqrt{\frac{\Lambda}{3\Gamma}  {+} \frac{\kappa [m S(t_0){+}W(t_0)]}{3\Gamma x^3}  {+}  \frac{\kappa \Psi^2_0 {\dot{\Phi}}^2_*(t_0)}{6\Gamma x^6}} \,.
\label{H}
\end{equation}
In terms of $x$ the key equation for the function $\Phi_*(x)$ is
\begin{equation}
H(x)\frac{d}{dx}{\Phi}_{*}(x) = \dot{\Phi}_*(t_0) x^{-4} \,,
\label{dxPhi}
\end{equation}
and we are ready for integration of this key equation.

\subsubsection{Solutions for the model with nonvanishing cosmological constant, $\Lambda \neq 0$}

When $\Lambda \neq 0$, we can rewrite the Hubble function (\ref{H}) as follows:
\begin{equation}
H(x) = \frac{H_{\infty}}{x^3} \sqrt{x^6 + \alpha x^3 + \beta} \,,
\label{H1}
\end{equation}
where we introduced the following auxiliary parameters:
\begin{equation}
H_{\infty} = \sqrt{\frac{\Lambda}{3\Gamma}} \,, \quad \alpha = \frac{\kappa [m S(t_0){+}W(t_0)]}{\Lambda} \,,
\label{parameters}
\end{equation}
$$
\beta=  \frac{\kappa \Psi^2_0 {\dot{\Phi}}^2_*(t_0)}{2\Lambda} \,.
$$
Formal integration of (\ref{dxPhi}) gives
\begin{equation}
3H_{\infty} \frac{ \Phi_*(x)}{\dot{\Phi}_*(t_0)} + const =
\label{solvePhi}
\end{equation}
$$
=  \frac{1}{\sqrt{\beta}} \log{\left[\frac{x^3}{2\sqrt{\beta(x^6+ \alpha x^3 + \beta)}+ \alpha x^3 + 2 \beta}\right]} \,.
$$
Since $x{=}1$ when $t{=}t_0$, the constant of integration can be easily found, yielding
\begin{equation}
\Phi_*(x) = \Phi_*(t_0) +
\label{solvePhi2}
\end{equation}
$$
+ \frac{\dot{\Phi}_*(t_0)}{3H_{\infty} \sqrt{\beta}} \log{\left\{\frac{x^3\left[2\sqrt{\beta(1+\alpha+\beta)} + \alpha + 2 \beta \right]}{\left[2\sqrt{\beta(x^6+\alpha x^3+\beta)}+ \alpha x^3+ 2\beta\right]}\right\}} \,.
$$
If we are interested to find $\Phi_*$ as a function of the cosmological time $t$,
we take the formula $\dot{x}=xH(x)$ and immediately  obtain
\begin{equation}
H_{\infty} (t-t_0) = \int_1^{\frac{a(t)}{a(t_0)}} \frac{x^2 dx}{\sqrt{x^6 + \alpha x^3 + \beta} }\,.
\label{t1}
\end{equation}
Direct integration in (\ref{t1}) gives
\begin{equation}
a(t) = \frac{a(t_0)}{2^{\frac13}} \left\{\sqrt{4\beta{-}\alpha^2}\sinh{\left[3H_{\infty}(t{-}t_*)\right]} {-} \alpha \right\}^{\frac13} \,.
\label{t2}
\end{equation}
We assume that $4\beta > \alpha^2$ providing $a(t)$ to be the real function of time; this is possible, when the positive cosmological constant is bigger than some critical value $\Lambda>\Lambda_C$, where
\begin{equation}
\Lambda_C = \frac{\kappa [m S(t_0){+}W(t_0)]^2}{2 \Psi^2_0 \dot{\Phi}^2_{*}(t_0)} \,.
\label{spec}
\end{equation}
Also, we introduce the auxiliary time moment $t_*$ as
\begin{equation}
t_* =  t_0 - \frac{1}{3H_{\infty}} {\rm arsh}\left(\frac{2+\alpha}{\sqrt{4\beta-\alpha^2}} \right) \,.
\label{t3}
\end{equation}
Asymptotic behavior of this field configuration relates to the de Sitter law
\begin{equation}
a(t \to \infty) \propto e^{H_{\infty} t} \,.
\label{t4}
\end{equation}
The function $\Phi_*$ tends asymptotically to the constant
\begin{equation}
\Phi_*(x \to \infty) = \Phi_*(t_0) +
\label{t5}
\end{equation}
$$
+\frac{\dot{\Phi}_*(t_0)}{3H_{\infty} \sqrt{\beta}} \log{\left\{\frac{\left[2\sqrt{\beta(1+\alpha {+}\beta)} {+} \alpha {+} 2 \beta \right]}{\left(2\sqrt{\beta }{+} \alpha \right)}\right\}} \,.
$$
Finally, it is interesting to calculate the acceleration parameter defined as
\begin{equation}
-q \equiv \frac{\ddot{a}}{aH^2} = 1 + \frac{x}{H} \frac{dH}{dx} \,.
\label{accel1}
\end{equation}
For the Hubble function (\ref{H1}) it has the form
\begin{equation}
-q \equiv \frac{x^6-\frac12 \alpha x^3 - 2\beta}{x^6 + \alpha x^3 +\beta} \,.
\label{accel2}
\end{equation}
Clearly, at the value of the reduced scale factor $x_{\rm T}$ there exists the cosmological transition point, for which $q(x_{\rm T})=0$ and thus the acceleration parameter changes the sign.
This point is calculated to be the following:
\begin{equation}
x_{\rm T} = \left(\frac{\alpha}{4} + \sqrt{\frac{\alpha^2}{16}+2\beta }\right)^{\frac13} \,.
\label{accel3}
\end{equation}
 Using the exact solution (\ref{t2}), one can rewrite this equality in terms of the cosmological time
\begin{equation}
t_{\rm T} = t_* + \frac{1}{3H_{\infty}}{\rm arsh}\left[\frac{3 + \sqrt{1+\frac{32\beta}{\alpha^2}}}{2\sqrt{\frac{4\beta}{\alpha^2}-1}} \right] \,.
\label{accel4}
\end{equation}
The function (\ref{accel2}) shows that in the model under consideration (we assume here that $\Lambda>0$ and thus $\alpha>0$ and $\beta>0$) the Universe expands with deceleration in the time interval $t_0<t<t_{\rm T}$. At $t>t_{\rm T}$ the expansion of the Universe becomes accelerated, and asymptotically $-q \to 1$. At present, we obtain the late-time accelerated expansion, as it should be.

The case, when $4\beta {=} \alpha^2$ has to be considered especially, since $t_*$ in (\ref{t3}) is now infinite.

\subsubsection{Special case $\beta {=} \frac14 \alpha^2$, $\Lambda \neq 0$ }

This special case can be realized if the cosmological constant takes the critical value, $\Lambda=\Lambda_C$.
For this special case the Hubble function (\ref{H1}) transforms into
\begin{equation}
H(x) = H_{\infty} \left(1 + \frac{\alpha}{2x^3}\right) \,,
\label{H11}
\end{equation}
and the scale factor has the form
\begin{equation}
a(t) = a(t_0) \left[\left(1+ \frac{\alpha}{2}\right)e^{3H_{\infty}(t-t_0)} - \frac{\alpha}{2}\right]^{\frac13} \,.
\label{H111}
\end{equation}
As for the guiding function, it is now of the form
\begin{equation}
\Phi_*(x) = \Phi_*(t_0) + \frac{2\dot{\Phi}_*(t_0)}{3H_{\infty} \alpha} \log{\left[\frac{x^3\left(1+ \frac{\alpha}{2}\right)}{\left(x^3+ \frac{\alpha}{2}\right)} \right]} \,.
\label{2solvePhi2}
\end{equation}
Asymptotic regime relates to the  de Sitter law $a \propto e^{H_{\infty}t}$, and the guiding function tends asymptotically to the constant
$\Phi_*(\infty) = \Phi_*(t_0) {+} \frac{2\dot{\Phi}_*(t_0)}{3H_{\infty} \alpha} \log{\left(1{+} \frac{\alpha}{2}\right)}$.

In this model the acceleration parameter can be reconstructed as follows
\begin{equation}
-q = \frac{x^3-\alpha}{x^3+\frac{\alpha}{2}} \,.
\label{accel11}
\end{equation}
The transition point is characterized by the reduced scale factor $x_{\rm T}=\alpha^{\frac13}$, and the corresponding time moment is
\begin{equation}
t_{\rm T} = t_0 + \frac{1}{3H_{\infty}}\log{\left(\frac{3\alpha}{2+\alpha}\right)} \,.
\label{accel14}
\end{equation}
When $\alpha>1$, we obtain that $t_{\rm T} > t_0$. Again, asymptotically the acceleration parameter tends to one, and we deal with the late-time acceleration, typical for the models with quasi-de Sitter asymptotes.

One can mention that, in both cases with $\Lambda \neq 0$ studied above the period of cosmological time $t_0<t<t_{\rm T}$ is characterized by the decelerated expansion, i.e., after the moment $t_0$ the inflationary stage can not be realized. This fact can be explained as follows. We consider the model in which at $t>t_0$ the spinor field is nonvanishing, i.e., massive fermions already filled the Universe. In other words, the moment of the phase transition, at which the fermions were born (we indicate it as $t_{\rm F}$) satisfies the inequality $t_{\rm F}<t_0$. This means, that the known inflection point on the graph of the scale factor evolution, which divides the epochs of the inflation and the first decelerated expansion, also belongs to the interval $t<t_0$. The model of such phase transition initiated by the dynamic aether is in progress, and we plan to extend the model correspondingly in the nearest future.

\subsubsection{Solutions for the model with vanishing cosmological constant, $\Lambda = 0$, and $m \neq 0$}

When $\Lambda=0$ and the spinor field is massive, $m \neq 0$, the Hubble function
\begin{equation}
H(x) = \frac{H_*}{x^3}\sqrt{x^{3}  +  \sigma}
\label{0L1}
\end{equation}
depends on two parameters
\begin{equation}
H_* = \sqrt{\frac{\kappa [m S(t_0){+}W(t_0)]}{3 \Gamma}} \,, \quad \sigma = \frac{\Psi^2_0 {\dot{\Phi}}^2_*(t_0)}{2[m S(t_0){+}W(t_0)]}  \,.
\label{0L2}
\end{equation}
Now we obtain
\begin{equation}
\Phi_*(x) = \Phi_*(t_0) +
\label{0L3}
\end{equation}
$$
 +\frac{\dot{\Phi}_*(t_0)}{3 H_* \sqrt{\sigma }} \log{\left[\frac{\left(\sqrt{1+\frac{x^3}{\sigma}}+1 \right)\left(\sqrt{1+\frac{1}{\sigma}}-1 \right)}{\left(\sqrt{1+\frac{x^3}{\sigma}}-1 \right)\left(\sqrt{1+\frac{1}{\sigma}}+1 \right)}\right]}\,.
$$
The corresponding dependence of the scale factor on the cosmological time $a(t)$ is given by
\begin{equation}
a(t) = a(t_0) \left\{\left[\frac32 H_*(t-t_0) + \sqrt{1 +  \sigma}\right]^2 -\sigma \right\}^{\frac13} \,.
\label{50L4}
\end{equation}
Asymptotic behavior of the scale factor is characterized by the power laws
\begin{equation}
a(t \to \infty) \to a(t_0) \left(\frac{3H_* t}{2}\right)^{\frac23} \,,
\label{0L5}
\end{equation}
and the asymptotic value of the guiding function is
\begin{equation}
\Phi_*(x \to \infty) = \Phi_*(t_0) {+}
 \frac{\dot{\Phi}_*(t_0)}{3\sqrt{\sigma}}
\log{\frac{\left(\sqrt{1{+}\frac{1}{\sigma}}{-}1 \right)}{\left(\sqrt{1{+}\frac{1}{\sigma}}{+}1 \right)}} \,.
\label{50L6}
\end{equation}
Since the parameter $\sigma$ is positive (see (\ref{0L2})), the acceleration parameter is now monotonic and negative:
\begin{equation}
-q = - \left[\frac{x^3+4\sigma}{2(x^3+\sigma)} \right] \,.
\label{accel21}
\end{equation}
This model can not explain the late-time accelerated expansion.

\subsubsection{Solutions for the model with  $\Lambda = 0$, $W(t_0)=0$ and $m = 0$}

In this special case the Hubble function (\ref{H})
\begin{equation}
H(x)= \gamma x^{-3} \,, \quad \gamma = \sqrt{\frac{\kappa \Psi^2_0 \dot{\Phi}^2_*(t_0)}{6\Gamma}} \,,
\label{001}
\end{equation}
tends to zero asymptotically, when $x \to \infty$. The corresponding scale factor
\begin{equation}
a(t)=a(t_0) \left[1+ 3\gamma(t-t_0) \right]^{\frac13}
\label{002}
\end{equation}
is of the power-law type, and the guiding function
\begin{equation}
\Phi_*(x) = \Phi_*(t_0) + \frac{\dot{\Phi}_*(t_0)}{\gamma} \log{x}
\label{003}
\end{equation}
has no finite asymptotic value. The acceleration parameter is negative $-q=-2$, the model is non-physical.

\section{Effective fermion mass induced by the coupling of the spinor field with axionic dark matter and dynamic aether}

\subsection{Reconstruction of the function $\Phi_*(S,P,\omega,\Omega)$}

We discuss here the simplest example of the function $\Phi_*(\rho)$, which depends on the argument
\begin{equation}
\rho = \tau_1 S^2 + \tau_2 \omega^2 + \tau_3(P^2 + \Omega^2) \,,
\label{rho}
\end{equation}
where $\tau_1$, $\tau_2$, $\tau_3$ are some model parameters. For the equilibrium configuration, we obtain from (\ref{good3}), (\ref{good4}), (\ref{good7}) that $\rho$ happens to be proportional to $x^{-6}$:
$$
\rho = \rho(t_0) x^{-6} \,,
$$
\begin{equation}
\rho(t_0) \equiv \left[\tau_1 S^2(t_0) {+} \tau_2 \omega^2(t_0) {+} \tau_3(P^2(t_0) {+} \Omega^2(t_0))\right] \,.
\label{rho2}
\end{equation}
Thus,  we can put the quantity
\begin{equation}
x= \left[\frac{\rho(t)}{\rho(t_0)} \right]^{-\frac16}
\label{xS}
\end{equation}
 into (\ref{solvePhi2}), (\ref{2solvePhi2}), or into (\ref{0L3}) obtaining the necessary function $\Phi_*(\rho)$.
Particularly, when $\tau_1=1$ and $\tau_2=\tau_3=0$, we see that $\rho = S^2$, and the behavior of the guiding function $\Phi_*$ is predetermined by the fermion number density $S$. When
$\tau_2=1$ and $\tau_1=\tau_3=0$, the interacting aether and spinor field regulate the state of the axionic dark matter via the function $\Phi_*(\omega)$.

\subsection{Oscillations of the effective spinor mass, induced by the axionic dark matter in the presence of the dynamic aether}

\subsubsection{Scheme of analysis}

We have found the guiding  function $\Phi_*(\rho)$ as the exact solution to the field equations of the model, for which the axionic dark matter is in the first Equilibrium state $\phi {=} \Phi_*$. Since in all the basic formulas (see, e.g., (\ref{keyGravity})) the spinor mass $m$ appears in the combination with the baryon energy density $W(t_0)$, as $[mS(t_0){+}W(t_0)]$, in this subsection only, for the sake of simplicity, we put $W(t_0)=0$. Now we consider the effective mass matrix of spinor particles based on the formula (\ref{2M}) with $\Phi_* {=} \Phi_*(\rho)$ for the pseudoscalar field $\phi$ near the Equilibrium ($\phi = \Phi_* + \xi$). We can estimate the induced mass $\mu \equiv <M>{-}m$ as follows:
\begin{equation}
\mu = -2 \left(\frac{T}{S}\right) \rho \frac{d\Phi_*}{d \rho} \approx  - \frac{m^2_A \Psi^2_0 }{3S(t_0)} \left(x^4 \frac{d\Phi_*}{d x}\right) \xi(t)
\,.
\label{MMM}
\end{equation}
The function $\xi(t)$ satisfies the linearized equation
\begin{equation}
\ddot{\xi} + 3 H \dot{\xi} +  m^2_A \xi =0 \,.
\label{xi}
\end{equation}
We assume that on the late-time expansion stage $H<<m_A$, and the third term in (\ref{xi}) seems to be much bigger than the second one. Thus, if we put the approximate solution
\begin{equation}
\xi(t) = \xi(t_1) \cos{m_A (t-t_1)} + \frac{\dot{\xi}(t_1)}{m_A} \sin{m_A (t-t_1)}
\label{xi2}
\end{equation}
into (\ref{MMM}) we can illustrate the idea about an axionically induced fermion mass oscillations. Clearly, the properties of the function ${\cal H} = x^4 \frac{d\Phi_*}{d x}$ predetermine the behavior of the effective mass variation $\mu$; below we calculate $\mu(t)$ for three cases described above.

\subsubsection{Induced mass $\mu$ in the case $\Lambda \neq 0$ and $\beta \neq \frac14 \alpha^2$}

In this general case the function ${\cal H}(x)$ has the form:
\begin{equation}
{\cal H}(x) = x^4\frac{d\Phi_*}{dx} = \frac{x^3}{\sqrt{x^6+ \alpha x^3 + \beta}} \,,
\label{ff}
\end{equation}
thus, taking into account (\ref{t2}) we obtain
$$
\mu(t) = - \xi(t)  \frac{m^2_A \Psi^2_0 \dot{\Phi}(t_0)}{3S(t_0)} \times
$$
\begin{equation}
\times \left\{\frac{\sinh{\left[3H_{\infty}(t{-}t_0)\right]} {-} \frac{\alpha}{\sqrt{4\beta{-}\alpha^2}}}{ \cosh{\left[3H_{\infty}(t{-}t_0)\right]}} \right\} \,.
\label{ffd}
\end{equation}
In the asymptotic regime ${\cal H}(x) \to 1$. If the spinor field is massless, $m=0$, we see that $\alpha=0$ and we deal with the formula
\begin{equation}
\mu(t) = - \xi(t)  \frac{m^2_A \Psi^2_0 \dot{\Phi}(t_0)}{3S(t_0)} \tanh{\left[3H_{\infty}(t{-}t_0)\right]} \,.
\label{fsd}
\end{equation}

\subsubsection{Induced mass $\mu$ in the case $\Lambda \neq 0$ and $\beta = \frac14 \alpha^2$}

In this special case, according to (\ref{2solvePhi2}) and (\ref{H111}), the spinor mass variation is given by the formula
$$
\mu = {-} \xi(t) \frac{m^2_A \Psi^2_0 \dot{\Phi}_*(t_0)}{3 S(t_0) H_{\infty}} \times
$$
\begin{equation}
\times \left[1{-}\frac{\kappa m S(t_0)}{\kappa m S(t_0){+} \Lambda} e^{{-}3H_{\infty}(t{-}t_0)}\right] \,.
\label{mu2}
\end{equation}
By the way, for the massless spinor field, $m=0$, and in the absence of a baryonic matter, this quantity behaves as $\mu(t) = - \mu^* \xi(t)$, with
\begin{equation}
\mu^*=\frac{m^2_A \Psi^2_0 \dot{\Phi}_*(t_0)}{3 S(t_0) H_{\infty}} \,.
\label{mu22}
\end{equation}

\subsubsection{The case $\Lambda =0$, $m\neq 0$}

The function ${\cal H}(x)$ calculated using the formula (\ref{0L3})
\begin{equation}
{\cal H}(x) = - \frac{\dot{\Phi}_*}{H_{*}} \cdot \frac{x^3}{\sqrt{x^3+\sigma}}
\,,
\label{mu15}
\end{equation}
shows that asymptotically, at $x \to \infty$ the spinor mass variation does not tend to constant, i.e., the model is instable.

\section{Discussion and conclusions}

Whenever we remember the grand event: the detection of neutrinos emitted due to the explosion of Supernova 1987A,  we try to imagine what could happen during the 168.000 light-years traveling of that neutrinos  from the Large Magellanic Cloud  to the Earth? The neutrinos born in such catastrophes could interact with  dark matter and dark energy, and one can try to find the fingerprints of the cosmic dark fluid in the data of neutrino observations. On the other hand, the neutrino flow from the Supernovae of the Type II (SN core-collapse) is predicted to be so huge, that the neutrinos can influence the state of dark matter in the source environment.

Keeping in mind this idea, we established the model of interaction between spinor field, axionic dark matter and dynamic aether. The central element of this model is the so-called guiding function $\Phi_*$, which is the detail of the potential of the pseudoscalar (axion) field (see (\ref{V})). Since this guiding function depends on the spinor field and dynamic aether via some scalars (we considered here four scalar (\ref{scalars})), we are faced with the nonlinear modifications of the equations for the vector, axion, spinor and gravitational fields. We have found the exact solutions to these modified field equations in the model with equilibrium axionic dark matter, in the framework of the isotropic homogeneous cosmology. For the model with nonvanishing cosmological constant the exact solutions for the guiding function are presented by (\ref{solvePhi2}) and (\ref{2solvePhi2}); when $\Lambda=0$, the corresponding exact solutions are of the form (\ref{0L3}) and (\ref{003}).

The influence of the dynamic aether reveals in the modification of the model working parameters by the factor $\Gamma$, which contains Jacobson's phenomenological constants (\ref{GammaC}). The constraint on the sum of two parameters $C_1{+}C_3$
has been obtained in 2017 due to the observation of the binary neutron star merger (the event encoded as GW170817 and GRB 170817A, see  \cite{170817}).
It was established that the ratio of the velocities of the gravitational and electromagnetic waves differs from one by the quantity about $10^{-15}$, and thus, the sum of the parameters $C_1{+}C_3$ is estimated as follows: $-6 \times 10^{-15}<C_1{+}C_3< 1.4 \times 10^{-15}$.
If we assume that $C_3 \simeq -C_1$, we have to write $\Gamma=1+ \frac32 C_2$ and to redefine correspondingly the parameter $H_{\infty}$.

Then we use the obtained exact solutions for $\Phi_*$ for finding of the so-called effective spinor mass (see (\ref{M}) for the effective spinor mass matrix, and (\ref{2M}) for the effective mass scalar). The results of calculations, which demonstrate the possibility of the induced spinor mass oscillations in the cosmological context, are presented by the formulas (\ref{ffd}), (\ref{fsd}), (\ref{mu2}). Clearly, the parameter $\mu_*$ defined by (\ref{mu22}) plays the role of effectiveness of oscillation production.  We have to emphasize that for finding of the guiding function $\Phi_*$ and of the induced spinor mass $\mu$ we do not calculate directly the components of the spinor field, but we have found the exact solutions for the spinor scalars (pseudoscalars) $S$, $P$, $\omega$ and $\Omega$, which are the arguments of the guiding function $\Phi_*$ (see the details in Subsection IIIA5).

The obtained results allow us to formulate the following three main conclusions.

\noindent
1. The cosmic spinor field influences the state of the axionic dark matter via the guiding function $\Phi_*$ (the argument of the  periodic potential of the axion field $V(\phi,\Phi_*)$), which is a time depending analog of the vacuum average value of the pseudoscalar field.

\noindent
2. Spinor particles (massive and massless) acquire effective masses due to the interaction with the axionic dark matter; variations of this effective mass are predetermined by the dynamics of the Universe expansion.

\noindent
3. The effective mass of the spinor field is proportional to the square of the axion mass, is inversely proportional to the square of the constant of the axion-photon coupling, depends on cosmological time,  and is exposed to oscillations with the frequency proportional to the axion mass $m_A$.

The final question, which we would like to touch, concerns the possibility to observe the predicted effect of axionically induced spinor mass oscillations. According to the recent report \cite{KAM}, the unique new Hyper-Kamiokande detector is announced to be able to monitor the atmospheric, supernova and solar neutrinos. We do not discuss the details and design of a new test, but its idea could be the following. The distance between the Sun and Earth is much less than the distance between the Supernova and Earth. If the detecting system is able to measure the time difference between a solar flare and a solar neutrino appearance in the terrestrial laboratory, it could give the difference between the velocities of the photons and neutrinos, and thus could be considered as a standard for the estimation of the neutrino mass in the model free of the dark matter influence. In contrast to this reference information, one can try to measure the corresponding time delay in the Supernova burst search; now the supernova neutrinos are assumed to be influenced by the axionic dark matter in the long trip to the Earth. Comparison of the corresponding estimation of the neutrino mass taking into account the axionically induced oscillations,  with the reference one, could help to test our hypothesis.

\vspace{3mm}

\acknowledgments{
This work was supported by the Russian Foundation for Basic Research (Grant N 20-52-05009)}

\end{document}